\documentstyle[epsf,12pt]{article}
\def\fnote#1#2{\begingroup\def\thefootnote{#1}\footnote{#2}\endgroup}

\begin{document}

\hfill{UTTG-04-01}

\vspace{36pt}

\begin{center}
{\large {\bf { Fluctuations in the Cosmic Microwave Background II:  $C_\ell$ at
Large and Small $\ell$}}}

\vspace{36pt}
Steven Weinberg\fnote{*}{Electronic address:
weinberg@physics.utexas.edu}\\
{\em Theory Group, Department of Physics, University of
Texas\\
Austin, TX, 78712}

\vspace{30pt}

\noindent
{\bf Abstract}
\end{center}

\noindent
General asymptotic formulas are given for the coefficient $C_\ell$ of the term
of multipole number $\ell$ in the temperature correlation function of the cosmic
microwave background, in terms of scalar and dipole form factors introduced in a
companion paper.  The formulas apply in two overlapping limits: for $\ell\gg 1$
and for $\ell d/d_A\ll 1$ (where $d_A$ is the angular diameter distance of the
surface of last scattering, and $d$ is a length, of the order of the acoustic
horizon at the time of last scattering,  that characterizes acoustic
oscillations before this time.) The frequently used approximation that $C_\ell$ receives its main contribution from wave numbers of order $\ell/d_A$ is found to 
be less accurate for the contribution of the Doppler effect than for
the Sachs--Wolfe effect and intrinsic temperature fluctuations.  For  $\ell
d/d_A\ll 1$ and $\ell\geq 2$, the growth of $C_\ell$ with $\ell$ is shown to be
affected by acoustic oscillation wave numbers of all scales.  The asymptotic
formulas are applied to a model of acoustic oscillations before the time of
last scattering, with results in reasonable agreement with more elaborate 
computer calculations.

\vfill

\pagebreak
\setcounter{page}{1}

\begin{center}
{\bf I. INTRODUCTION }
\end{center}

A companion paper[1] has shown how to express the temperature fluctuation in the
cosmic microwave background in any direction as an integral involving scalar and
dipole form factors $F(k)$ and $G(k)$, which characterize acoustic oscillations
before the time of last scattering.     In the present paper we derive
asymptotic formulas for the strength $C_\ell$ of fluctuations at  multipole
number $\ell$ for form factors of arbitary functional form.  After outlining our
assumptions and reviewing some generalities in Section II, our general result in
the limit of $\ell\gg 1$ [Eq.~(26)] is derived in Section III.  In this limit
$\ell(\ell+1) C_\ell$ depends on $\ell$ and the angular diameter distance $d_A$
at the
time
of last scattering  only through the ratio $\ell/d_A$.  (This is why the heights
of the Doppler peaks do not depend on parameters like the cosmological constant
that affect $d_A$ but not the form factors.)  Our result in
the limit $\ell d/d_A\ll 1$ [Eq.~(43)] is derived in Section IV.  (Here $d$ is
some length characterizing acoustic oscillations, such as
the acoustic horizon distance $d_H$ at the time of last scattering).  These
ranges of $\ell$
overlap because $d_A\gg d$.

Even without a detailed calculation of the form factors, these results have a
moral for the physical interpretation of measurements of $C_\ell$.  It is common
to interpret these measurements by supposing that $C_\ell$ arises mostly from
fluctuations of wave number $k\simeq \ell /d_A$.  Eq.~(27) shows that this is a
fair approximation for the contribution of the scalar form factor $F(k)$, which
represents the Sachs--Wolfe effect and intrinsic temperature fluctuations;
$C_\ell$ receives no contribution from $F(k)$ with $k<\ell/d_A$, and the
contribution from $k\gg\ell/d_A$ is suppressed by a factor $\beta^{-2}(\beta^2-
1)^{-1/2}$, where $\beta\equiv kd_A/\ell$.  In particular, a peak in the
magnitude of the scalar form factor $F(k)$ at some wave number $k_1$ (like the
peak found in the simple model studied in reference [1] at $k=\pi/d_H$) will
show up in $\ell(\ell+1)C_\ell$ at a value of $\ell$ less than but close to $k_1
d_A$.  For instance, we will see in Section V that the peak in $|F(k)|$ at
$k=\pi/d_H$ produces a peak in $\ell(\ell+1)C_\ell$ at $\ell\simeq 2.6 d_A/d_H$
rather than at $\pi d_A/d_H$.  But Eq.~(27) also shows that this
interpretation of $C_\ell $ is much less useful for the contribution of the
vector form factor $G(k)$, which arises from the Doppler effect;
$C_\ell$ also receives no contribution from $G(k)$ with $k<\ell/d_A$, but
instead of the contribution from $k\simeq\ell/d_A$ being enhanced by a factor
$(\beta^2-1)^{-1/2}$, it is {\em suppressed} by a factor $(\beta^2-1)^{1/2}$.
Indeed, we will see in Section V that for sufficiently small baryon number the
peak in $G(k)$ at $k=\pi/2d_H$ found in the simple model of reference [1] does
show up as a peak in $\ell(\ell+1)C_\ell$, but at $\ell\simeq .45 d_A/d_H$, much
less than $(\pi/2)d_A/d_H$.
Furthermore, the behavior of $\ell(\ell+1)C_\ell$ for $\ell d/d_A$  near zero
depends
on the values of $F(k)$ and $G(k)$ for all $k$.  This points up the value of
observations that can measure the correlation function of
temperature
fluctuations directly, as a supplement to measurements of $C_\ell$.

The results obtained in Sections III and IV are used in Section V to calculate
$C_\ell$ for the approximate form factors calculated in reference 1.
In agreement with what  is found in more accurate computer calculations, the
position $\ell_1$ of the first Doppler peak is not a sensitive function of the
baryon density parameter $\Omega_Bh^2$.  On the other hand, we find that the
ratio of the value of $\ell(\ell+1)C_\ell$ at the first
Doppler peak to its value at
$\ell\ll d_A/d_H$ is a sensitive indicator of the value of $\Omega_Bh^2$.

\begin{center}
{\bf II. GENERALITIES}
\end{center}

The companion paper[1] shows that, in very general models (but assuming only
compressional normal modes, with no gravitational radiation), the fractional
variation from the mean of the cosmic microwave background temperature observed
in a direction $\hat{n}$ takes the general form
\begin{equation}
\frac{\Delta T(\hat{n})}{T}
=\int d^3k\,\epsilon_{\bf k}\,e^{id_A\hat{n}\cdot{\bf
k}}\,\left[F(k)+i\,\hat{n}\cdot\hat{k}\,G(k)\right]\;,
\end{equation}
aside from effects arising from late times, which chiefly affect the
coefficients $C_\ell$ for relatively small $\ell$.
Here $d_A$ is the angular diameter distance of the surface of
last scattering
\begin{equation}
d_A=\frac{1}{\Omega_C^{1/2}H_0(1+z_L)}\sinh\left[\Omega_C^{1/2}\int_{\frac{1}{1+
z_L}}^1\frac{dx}
{\sqrt{\Omega_\Lambda x^4+\Omega_Cx^2+\Omega_Mx}}\right]\;,
\end{equation}
where $\Omega_C\equiv1-\Omega_\Lambda-\Omega_M$, and $\Omega_\Lambda$ and
$\Omega_M$ are the present ratios of the energy densities of the vacuum and
matter to the critical density $3H_0^2/8\pi G$.  (If the vacuum energy were to
change with time, as in theories of quintessence, then the formula for $d_A$
would need modification, but there would be essentially no change in the other
ingredients in Eq.~(1), as long as the quintessence energy density makes a
negligible contribution to the total energy density at and before the time of
last scattering.)  Also, ${\bf k}^2\epsilon_{\bf k}$ is proportional
(with a ${\bf k}$-independent proportionality coefficient) to the Fourier
transform of the fractional
perturbation in the energy density early in the radiation-dominated era.
The average\fnote{**}{The average here is over an ensemble of possible
fluctuations.  Using Eq.~(3) to analyze the particular element of this sample
observed in our universe relies on ergodic arguments, which are not exact except
in the limit $\ell\rightarrow \infty$.  However, corrections are manageable[2]
even for small $\ell$.} of the product of two $\epsilon$s is assumed to satisfy
the
conditions of statistical homogeneity and isotropy:
\begin{equation}
\left\langle \epsilon_{\bf k}\,\epsilon_{\bf k'}\right\rangle
=\delta^3({\bf k}+{\bf k'})\,{\cal P}(k)
\end{equation}
with $k\equiv |{\bf k}|$.  The
power spectral function ${\cal P}(k)$ is real and positive.  Where a specific
expression for ${\cal P}(k)$ is needed, we will use the `scale-invariant' (or
$n=1$) Harrison--Zel'dovich form suggested by theories of new inflation:
\begin{equation}
{\cal P}(k)=B\,k^{-3}\;,
\end{equation}
with $B$ a constant that must be taken from observations of the cosmic microwave
background or condensed object mass distributions, or from detailed theories of
inflation.

The form factors $F(k)$ and $G(k)$ characterize acoustic oscillations, with
$F(k)$ arising from the Sachs--Wolfe effect and intrinsic temperature
fluctuations, and $G(k)$ arising from the Doppler effect.  For instance, they
are calculated in  reference 1 in the approximation that perturbations in the
gravitational field at and before the time of last scattering arise entirely
from perturbations in the density of cold dark matter.  For very small wave
numbers the form factors are
\begin{eqnarray}
F(k)&\rightarrow & 1-3k^2t_L^2/2-3[-\xi^{-
1}+\xi^{-2}\ln(1+\xi)]k^4t_L^4/4+\dots\;,\\
G(k)&\rightarrow & 3kt_L-3k^3t_L^3/2(1+\xi)+\dots\;,
\end{eqnarray}
while for wave numbers large enough to allow the use of the WKB approximation,
i. e.,
\begin{equation}
kt_L> \xi
\end{equation}
the form factors are
\begin{equation}
F(k)= (1-2\xi/k^2t_L^2)^{-1}\left[-3\xi+2\xi/k^2t_L^2+(1+\xi)^{-1/4}e^{-
k^2d_D}\cos(kd_H)\right]\;,
\end{equation}
 and
\begin{equation}
G(k)=\sqrt{3}\,(1-2\xi/k^2t_L^2)^{-1} (1+\xi)^{-3/4} e^{-
k^2d_D}\sin(kd_H)\;.
\end{equation}
Here $t_L$ is the time of last scattering; $\xi$ is $3/4$ the ratio of the
baryon to photon energy densities at this time:
\begin{equation}
\xi=\left(\frac{3\rho_B}{4\rho_\gamma}\right)_{t=t_L}=27\,\Omega_Bh^2\;;
\end{equation}
$d_H$ is the acoustic horizon size at this time,
and $d_D$ is a damping length, given by Eq.~(48).
These formulas for the form factors are mentioned at this point only for
illustration; we
will be working here with general form factors $F(k)$ and $G(k)$, and will not
make use of the specific formulas (5)--(10) until Section V.
But we will assume throughout that any lengths $d$ that (like $d_H$ and
$d_D$ in Eqs.~(8) and (9)) characterize the $k$-dependence of the form
factors are much smaller than the angular diameter
distance $d_A$ of last scattering.  This is a good approximation: for instance,
if
the ratios of matter and vacuum energy densities to the critical density
have the present values $\Omega_M=0.3$ and $\Omega_\Lambda=0.7$, then
$d_A/d_H$ runs from 91.7 to 79.7 for values of $\Omega_B h^2$ running from zero
to 0.03,  and $d_D$ is smaller than $d_H$, independent of the value of
$H_0$.

It is usual to employ the well-known expansion of a plane wave in Legendre
polynomials, and write Eq.~(1) as
\begin{eqnarray}
\frac{\Delta T(\hat{n})}{T}&=&\sum_{\ell=0}^\infty  (2\ell+1)\,i^\ell
\int d^3k\;\epsilon_{\bf
k}\,P_\ell\left(\hat{n}\cdot\hat{k}\right)\Bigg[j_\ell\left(kd_A\right)\,F(k)
\nonumber\\&&\qquad+ j'_\ell\left(kd_A\right)\,G(k) \Bigg]\;.
\end{eqnarray}
Using Eq.~(3) and the orthogonality property of Legendre polynomials
\begin{equation}
\int d\Omega_{\hat{k}} \,P_\ell\left(\hat{n}\cdot\hat{k}\right)\,
P_{\ell'}\left(\hat{n}'\cdot\hat{k}\right)=\left(\frac{4\pi}{2\ell+1}\right)
\delta_{\ell\,\ell'}P_\ell\left(\hat{n}\cdot\hat{n}'\right)\;,
\end{equation}
one finds that
\begin{equation}
\left\langle \frac{\Delta T(\hat{n})}{T}\; \frac{\Delta T(\hat{n'})}{T}
\right\rangle=\sum_{\ell =0}^\infty
\left(\frac{2\ell+1}{4\pi}\right)\,C_\ell\,P_\ell\left(
\hat{n}\cdot\hat{n}'\right)\;,
\end{equation}
with the conventional coefficient $C_\ell$ taking the value
\begin{equation}
C_\ell = 16\pi^2\int_0^\infty {\cal
P}(k)\,k^2\,dk\,\Big[j_\ell\left(kd_A\right)\,F(k) +
j'_\ell\left(kd_A\right)\,G(k) \Big]^2\;.
\end{equation}
This familiar formula is adequate for numerical calculation of $C_\ell$, but it
hides the
essential qualitative aspects of the dependence of $C_\ell$ on $\ell$: that
$C_\ell$ for $\ell\gg 1$ depends on the ratio $\ell/d_A$, and that
$\ell(\ell+1)C_\ell$ approaches a constant for sufficiently small values of this
ratio,
whether $\ell$ itself is large or small.  To obtain these results, we must now
distinguish between the two  cases $\ell\gg 1$ and $\ell\ll d_A/d$ (but
$\ell\geq 2$),
where $d$ is a typical length characteristic of the form-factors $F(k)$ and
$G(k)$.   These two cases overlap because, as remarked above, $d_A$ is much
larger than $d$.

\begin{center}
{\bf III. LARGE $\ell$ }
\end{center}

The usual way of obtaining the contribution of the scalar form factor to
$C_\ell$ for large $\ell$ is to note
that the integral (14) receives its largest contribution when the argument of
the
spherical Bessel function is of order $\ell$, in which case we can use the
approximation that, for $\ell\rightarrow\infty$,
\begin{equation}
j_\ell(z)\rightarrow \left\{\begin{array}{ll} 0 &\; z<\nu \\ z^{-1/2}(z^2-
\nu^2)^{-1/4}\cos\left(\sqrt{z^2-\nu^2}-\nu\arccos(\nu/z)-\frac{\pi}{4}\right)
 &\; z>\nu\;,\end{array}\right.~~~~~~~
\end{equation}
where  $z/\nu$ is held fixed at a value $\neq 1$, with $\nu\equiv \ell+1/2$.
The procedure is straightorward for the $F^2$ terms in Eq.~(14), but for the
$FG$
and $G^2$ terms involving the Doppler effect we run into a difficulty:
differentiating the factor $(z^2-\nu^2)^{-1/4}$ in Eq.~(15) yields larger
negative powers of
$z^2-\nu^2$ that introduce divergences from the part of the integral in Eq.~(14)
near the lower bound $k=\nu/d_A$.  These infrared divergences are spurious,
because
the asymptotic formula (15) breaks down if we let $z$ and
$\nu$ go to infinity  in such a way that $z/\nu\rightarrow 1$.  This problem can
be dealt with by switching to a
different asymptotic limit[3] for $k$ near $\nu/d_A$.  Here we will use a
different
method[4] which avoids the delicate problem of the asymptotic behavior of
$j_\ell(z)$ and
$j'_\ell(z)$ for $z$ near $\nu$.

We return to Eq.~(1), and use Eq.~(3) to put the correlation function
of observed temperature fluctuations in the form
\begin{eqnarray}
&&\left\langle \frac{\Delta T(\hat{n})}{T}\; \frac{\Delta T(\hat{n'})}{T}
\right\rangle =\int d^3k\,{\cal P}(k)\,\exp\left(id_A{\bf k}\cdot(\hat{n}-
\hat{n}')\right)\,
\left[F^2(k) \right.\nonumber\\&&\qquad\qquad\left.+i\hat{k}\cdot(\hat{n}-
\hat{n}')F(k)\,G(k)+(\hat{k}\cdot\hat{n})(\hat{k}\cdot\hat{n}')\,G^2(k)\right]\;
.
\end{eqnarray}
The integral over the direction of ${\bf k}$ is easy, and gives the correlation
function
\begin{eqnarray}
&&\left\langle \frac{\Delta T(\hat{n})}{T}\; \frac{\Delta T(\hat{n'})}{T}
\right\rangle =4\pi\int_0^\infty k^2\,dk\,{\cal P}(k)\,
\left[F^2(k)+ F(k)\,G(k)\,\frac{\partial}{\partial (d_Ak)}
\right.\nonumber\\&&\quad\left.+
\frac{1}{2}G^2(k)\left(1+\frac{\theta^4}{4}+\left(\frac{1}{\theta^2}-
\frac{1}{2}+\frac{3\theta^2}{4}\right)\frac{\partial^2}{\partial
(d_Ak)^2}\right)\right]\frac{\sin (d_Ak\theta)}{d_Ak\theta}\;,~~~~
\end{eqnarray}
where $\theta\equiv |\hat{n}-\hat{n}'|$.  (This formula may prove useful in
analyzing observations that give the correlation function
directly, rather than in terms of $C_\ell$.)  The amplitude $C_\ell$ is defined
as
the integral
\begin{equation}
C_\ell= 2\pi\int_{-1}^{+1} P_\ell(\mu)\left\langle \frac{\Delta T(\hat{n})}{T}\;
\frac{\Delta T(\hat{n'})}{T} \right\rangle \,d\mu\;,
\end{equation}
where $\mu\equiv \hat{n}\cdot\hat{n}'=1-\theta^2/2$.  For large $\ell$ the
Legendre polynomial $P_\ell(\mu)$ oscillates rapidly for $\theta\gg 1/\ell$, so
the integral is dominated by values of $\theta$ of order $1/\ell$, in which case
we can use the well-known limiting expression $P_\ell(\mu)\rightarrow
J_0(\ell\theta)$, and
write
\begin{eqnarray}
C_\ell&\rightarrow &8\pi^2 \int_0^\infty k^2\,dk\, {\cal P}(k)\int_0^2
J_0(\ell\theta)\,\theta\,d\theta \left[F^2(k)+
F(k)\,G(k)\,\frac{\partial}{\partial (d_Ak)} \right.\nonumber\\&&\left.+
\frac{1}{2}G^2(k)\left(1+\frac{\theta^4}{4}+\left(\frac{1}{\theta^2}-
\frac{1}{2}+\frac{3\theta^2}{4}\right)\frac{\partial^2}{\partial
(d_Ak)^2}\right)\right]\frac{\sin (d_Ak\theta)}{d_Ak\theta}\;.~~~~~~
\end{eqnarray}
The integral over $k$ is dominated by values for which $kd_A\theta$ is of order
unity, so the derivative $\partial/\partial (d_Ak)$ is effectively of order
$\theta\approx 1/\ell$.
Thus to leading order in $1/\ell$, Eq.~(19) may be simplified to
\begin{eqnarray}
C_\ell&\rightarrow &8\pi^2 \int_0^\infty k^2\,dk\, {\cal P}(k)\int_0^2
J_0(\ell\theta)\,\theta\,d\theta \nonumber\\&& \times \left[F^2(k) +
\frac{1}{2}G^2(k)\left(1 +\frac{1}{\theta^2}\frac{\partial^2}{\partial
(d_Ak)^2}\right)\right]\frac{\sin (d_Ak\theta)}{d_Ak\theta}\;.~~~~~~
\end{eqnarray}
Introducing a new variable $s\equiv \ell\theta$ and changing the upper limit on
the $s$-integral from $2\ell$ to infinity, we may write this as
\begin{eqnarray}
C_\ell&\rightarrow &\frac{8\pi^2 }{\ell^2}\int_0^\infty k^2\,dk\, {\cal
P}(k)\int_0^\infty J_0(s)\,s\,ds \nonumber\\&& \times \left[F^2(k) +
\frac{1}{2}G^2(k)\left(1 +\frac{\partial^2}{\partial
(d_Aks/\ell)^2}\right)\right]\frac{\sin (d_Aks/\ell)}{(d_Aks/\ell)}\;.~~~~~~
\end{eqnarray}
The integral over $s$ is easy for the $F^2$ term; we need only use the
formula[5]:
\begin{equation}
\int_0^\infty J_0(s)\sin(\beta s)\,ds=\left\{\begin{array}{ll} 0 &\qquad \beta<1
\\
(\beta^2-1)^{-1/2} & \qquad\beta>1\;,\end{array}\right.
\end{equation}
where here $\beta=d_Ak/\ell$.
The integral of the $G^2$ term takes a little more work.  We use the formula
$(1+d^2/dx^2)\sin x/x=-(2/x)d/dx (\sin x/x)$ and do the remaining integral by
parts,
so that
\begin{eqnarray}
&&\int_0^\infty J_0(s) \,s \,\left[1+\frac{\partial^2}{\partial (\beta
s)^2}\right]\frac{\sin (\beta s)}{\beta s} \,ds
=-\frac{2}{\beta^2}\int_0^\infty J_0(s) \frac{\partial}{\partial  s}\frac{\sin
(\beta s)}{s}\,ds
\nonumber\\&& =\frac{2}{\beta^2}-\frac{1}{\beta^3}\int_0^\infty
\Big(J_2(s)+J_0(s)\Big)\,\sin(\beta s)\,ds\;.
\end{eqnarray}
Here we also need the formula[5]
\begin{equation}
\int_0^\infty J_2(s)\sin(\beta s)\,ds=\left\{\begin{array}{ll} 2\beta &
\qquad\beta<1 \\
-(\beta^2-1)^{-1/2}(\beta+\sqrt{\beta^2-1})^{-1} &\qquad
\beta>1\end{array}\;,\right.
\end{equation}
so that
\begin{equation}
\int_0^\infty J_0(s) \,s \,\left[1+\frac{\partial^2}{\partial (\beta
s)^2}\right]\frac{\sin (\beta s)}{\beta s} \,ds=\left\{\begin{array}{ll}
0 & \qquad\beta<1 \\ 2\beta^{-3}\sqrt{\beta^2-1} &
\qquad\beta>1\;.\end{array}\right.
\end{equation}
Using Eqs.~(22) and (25) in Eq.~(21) then gives our final general formula for
$C_\ell$ at large $\ell$:
\begin{equation}
C_\ell \rightarrow \frac{8\pi^2\ell }{d_A^3}\int_1^\infty d\beta\, {\cal
P}(\ell\beta/d_A) \left[\frac{\beta F^2(\ell\beta/d_A )}{\sqrt{\beta^2-1}} +
\frac{\sqrt{\beta^2-1}\,G^2(\ell\beta/d_A )}{\beta} \right]\;.
\end{equation}
Note that $\ell^2 C_\ell$ depends on $\ell$ and $d_A$ only through its
dependence on the ratio $\ell/d_A$.

For instance, if take the power spectral function to have the
scale-invariant  form  ${\cal P}(k)=Bk^{-3}$, then for $\ell\gg 1$
\begin{equation}
\ell(\ell+1) C_\ell \rightarrow 8\pi^2 B\int_1^\infty d\beta\,  \left[\frac{
F^2(\ell\beta/d_A )}{\beta^2\sqrt{\beta^2-1}} + \frac{\sqrt{\beta^2-
1}\,G^2(\ell\beta/d_A )}{\beta^4} \right]\;.
\end{equation}
(We have taken advantage of the fact that here we are considering $\ell\gg 1$ to
change a factor $\ell^2$ to $\ell(\ell+1)$, in order to facilitate comparison
with the results of the next section.)
The rapid fall-off of the coefficient of $F^2$ for $\beta>1$ suggests that the
contribution of the scalar form factor $F$ to $C_\ell$ is
dominated by wave numbers close to $d_A/\ell$, as is usually assumed.  On the
other hand, the contribution of the dipole form factor $G(k)$ for wave numbers
immediately above $d_A/\ell$ is actually suppressed by the factor
$\sqrt{\beta^2-1}$ in the second term of Eq.~(27).

\begin{center}
{\bf IV. SMALL $\ell\, d /d_A$ }
\end{center}

Here we will adopt the `$n=1$' scale-invariant spectrum  ${\cal P}(k)\simeq
B k^{-3}$ from the beginning, so that the general formula Eq.~(14) becomes
\begin{equation}
C_\ell=16\pi^2 B\int_0^\infty \left[j_\ell(s)\,F\left(\frac{s}{d_A}\right)
+ j'_\ell(s)\,G\left(\frac{s}{d_A}\right)\right]^2\frac{ds}{s}\;.
\end{equation}
To generate a series for $\ell(\ell+1)C_\ell$ in powers of
$\ell/d_A$ we
expand the form factors in power series:
\begin{equation}
F(k)=F_0+F_2\,k^2+\cdots\;,~~~~~~~~G(k)=G_1k+G_3k^3+\cdots\;.
\end{equation}
(The power series for $F$ and $G$ must be respectively even and odd in $k$, in
order that the integrand in the temperature fluctuation (1) should be analytic
in the three-vector ${\bf k}$ at ${\bf k}=0$.)
The leading term in $C_\ell$ is well known;
using a standard formula[6]:
\begin{equation}
\int_0^\infty j_\ell^2(s)\,s^{m-1}ds= \frac{2^{m-3}\pi \Gamma(2-
m)\,\Gamma\left(\ell+\frac{m}{2}\right)}{\Gamma^2\left(\frac{3-
m}{2}\right)\,\Gamma\left(\ell+2-\frac{m}{2}\right)}\;,
\end{equation}
we find the term in Eq.~(28) of zeroth order in $1/d_A$:
\begin{equation}
C^{(0)}_\ell=\frac{8\pi^2 B F_0^2}{\ell(\ell+1)}\;.
\end{equation}
There is no difficulty in also calculating the term in Eq.~(28) of first order
in
$1/d_A$:
\begin{equation}
C^{(1)}_\ell=\left(\frac{32\pi^2 \,B F_0 G_1}{d_A}\right)\,\int_0^\infty
j_\ell(s)\,j'_\ell(s)\,ds= \left(\frac{16\pi^2 \,B F_0 G_1}{d_A}\right)\,
\left[j^2_\ell (s)\right]_{0}^\infty=0\;.
\end{equation}
But we run into trouble in calculating the term of second order in $1/d_A$.
The second derivative of $C_\ell$ with respect to $1/d_A$ is
\begin{eqnarray}
\frac{d^2 C_\ell}{d\,(1/d_A)^2}&=&
16\pi^2 B \int_0^\infty\left\{j_\ell^2(s) {F^2}''(s/d_A)
+ {j'}_\ell^2(s)
{G^2}''(s/d_A)\right.\nonumber\\&&\left.+2j_\ell(s)j'_\ell(s)\Big[F(s/d_A)G(s/d
_A)\Big]''
\right\}\,s\,ds\;.
\end{eqnarray}
The $j_\ell j'_\ell$ term doesn't contribute to the part of $C_\ell$ of second
order in $1/d_A$, because $F(k)G(k)$ contains only odd powers of $k$.  To
calculate the contribution of the $ {j'}^2_\ell$ term, we need to supplement
Eq.~(30) with the additional formula:\fnote{\dagger }{ This formula was obtained
by
using the Bessel differential equation to
show that ${j'_\ell}^2(z)=(1-\ell(\ell+1)/z^2)j_\ell^2(z)+(zj_\ell^2(z))''/2z$,
and then using Eq.~(30) with two integrations by parts.}
\begin{eqnarray}
&&\int_0^\infty {j'_\ell}^2(s)\,s^{m-1}ds= \frac{2^{m-3}\pi \Gamma(2-
m)\,\Gamma\left(\ell+\frac{m}{2}\right)}{\Gamma^2\left(\frac{3-
m}{2}\right)\,\Gamma\left(\ell+2-\frac{m}{2}\right)}\nonumber\\&&\quad
\times\,\left[1+\frac{(m-3)(m-2) \Big((m-2)(m-3)-2\ell(\ell+1)\Big)}{2(3-
m)^2\left(\ell+\frac{m}{2}-1\right)\left(\ell-
\frac{m}{2}+2\right)}\right]\;.~~~~~~~
\end{eqnarray}
The second derivative (33) is divergent at $1/d_A=0$, as shown by the factors
$\Gamma(2-m)$ in Eqs.~(30) and (34), which become infinite for $m=2$.  Of
course, there is no infinity in $C_\ell$; it is simply not analytic in $1/d_A$
at $1/d_A=0$.

We can deal with this problem by a method similar to the dimensional
regularization technique used in quantum field theory[7].  We treat $m$ as a
complex variable that approaches $m=2$.  In this limit, Eqs.~(30) and (34) give
\begin{equation}
\int_0^\infty {j_\ell}^2(s)\,s^{m-1}ds\rightarrow -\frac{1}{2}\left[\frac{1}{m-
2}+\sum_{r=1}^\ell \frac{1}{r}-C+\ln 2-D\right]\;,
\end{equation}
\begin{equation}
\int_0^\infty {j'_\ell}^2(s)\,s^{m-1}ds\rightarrow  -
\frac{1}{2}\left[\frac{1}{m-2}+\sum_{r=1}^\ell \frac{1}{r}-C+\ln 2-D+1\right]\;,
\end{equation}
where $C$ is the Euler constant $C\equiv -\Gamma'(1)=0.57722 $, and $D\equiv -
\Gamma'(1/2)/\Gamma(1/2)=1.96351$.
The important point here is that the parts of the integrals (35) and (36) that
are divergent at $m=2$ are independent of $\ell$, and so also is the part of
$C_\ell$ that is non-analytic in $1/d_A$ at $1/d_A=0$.  Using Eqs.~(29), (35)
and (36) in Eq.~(33) thus gives
the part of $C_\ell$ that is of second order in   $1/d_A$ as
\begin{equation}
C_\ell^{(2)}=-8\pi^2 B \,d_A^{-2}\,\Big(2F_0F_2+G_1^2\Big)\sum_{r=1}^\ell
\frac{1}{r}\;\;+\;\;\ell\!-\!{\rm independent\;terms}\;.
\end{equation}

We can check the consistency of these results and calculate the
$\ell$-independent terms here by using our previous result
(27) in the case where  $\ell$ is large {\em and} $ \ell d /d_A$ is small,
where $d$ is whatever length characterizes the $k$-dependence of the form
factors.  The term in Eq.~(27) of zeroth order in $\ell d /d_A$ is
\begin{equation}
\ell(\ell+1) C_\ell\rightarrow 8\pi^2
BF_0^2\int_1^\infty\frac{d\beta}{\beta^2\sqrt{\beta^2-1}}=8\pi^2 BF_0^2\;,
\end{equation}
in agreement with Eq.~(31).  Also, Eq.~(27) has no terms of first order in
$1/d_A$, in agreement with Eq.~(32).
To calculate the terms in Eq.~(27) of second order in $1/d_A$, we
express $F^2(k)$ and $G^2(k)$ in terms of cosine transforms
\begin{equation}
F^2(k)=F_0^2+\int_0^\infty da\;f(a)\,\Big(1-\cos(ka)\Big)\;,~~~~~~~
G^2(k)=\int_0^\infty da\;g(a)\,\Big(1-\cos(ka)\Big)\;.
\end{equation}
Then for $\ell\gg 1$ {\it and} $\ell d/d_A\ll 1$, Eq.~(27) gives
\begin{equation}
C^{(2)}_\ell\rightarrow -
\frac{8\pi^2B}{d_A^2}\left[\left(2F_0F_2+G_1^2\right)\,\left(\ln\left(\frac{\ell
\bar{d}}{2d_A}\right)+C-\frac{3}{2}\right)+G_1^2\right]\;,
\end{equation}
where $\bar{d}$ is a typical value of the variable $a$ in the cosine transforms
(39):
\begin{equation}
\ln\bar{d}\equiv \frac{\int_0^\infty \left[f(a)+g(a)\right]\,a^2\,\ln
a\,da}{\int_0^\infty \left[f(a)+g(a)\right]\,a^2\,da }
\end{equation}
Eq.~(40)  agrees with the limit of Eq.~(37) for large $\ell$, because in this
limit $\sum_1^\ell 1/r\rightarrow\ln\ell+C$, and now fixes the
$\ell$-independent terms in Eq.~(37) so that, for any $\ell$ with $\ell d/d_A\ll
1$,
\begin{equation}
C^{(2)}_\ell= -
\frac{8\pi^2B}{d_A^2}\left[\left(2F_0F_2+G_1^2\right)\,\left(\ln\left(\frac{\bar
{d}}{2d_A}\right)+\sum_{r=1}^\ell\frac{1}{r}-\frac{3}{2}\right)+G_1^2\right]\;.
\end{equation}

Putting together Eqs.~(31), (32), and (42) gives our final formula for $C_\ell$
in the case $\ell d/d_A\ll 1$ and $\ell\geq 2$:
\begin{equation}
\ell(\ell+1)C_\ell=8\pi^2BF_0^2\left\{1-\frac{
\ell(\ell+1)}{d_A^2}\left[d^2\,\left(\ln\left(\frac{\bar{d}}{2d_A}\right)
+\sum_{r=1}^\ell\frac{1}{r}\right)- d'^2\right]+\dots\right\}\;,
\end{equation}
where now we introduce a pair of  characteristic lengths:
\begin{equation}
d^2\equiv \frac{2F_0F_2+G_1^2}{F^2_0}\;,~~~~~~ d'^2\equiv
\frac{3F_0F_2+\frac{1}{2}G_1^2}{F^2_0}
\;.
\end{equation}
The logarithm in Eq.~(43) is large and negative, so
$\ell(\ell+1)C_\ell$ will  increase or decrease  with $\ell$ for sufficiently
small
$\ell$ according as $d^2>0$ or $d^2<0$.  (Taken literally, Eq.~(43) would
suggest that this behavior is reversed when the sum over $r$ becomes large
enough to cancel the logarithm, but this is at $\ell\simeq 2e^{-C}d_A/\bar{d}$,
which is large enough to invalidate the approximations that led to Eq.~(43).)
Note that, while $d$ and $d'$ depend only on the behavior of the form factors
near zero wave number, the length $\bar{d}$ given by Eq.~(41) depends on the
behavior of the form factors at all wave numbers.  Consequently,  although the
{\em value} of $C_\ell$ at low $\ell$ depends only on the form factors at $k=0$,
somewhat surprisingly the {\em growth} of $C_\ell$ for small $\ell$
depends on the form factors at all wave numbers.

\begin{center}
{\bf V. APPLICATION}
\end{center}

To illustrate the use of the asymptotic formulas obtained here, we will now
apply them to the simplified model described in reference 1: the universe before
last scattering consisting of pressureless cold dark matter and a
photon-nucleon-electron plasma; no gravitational radiation; and negligible
contributions of the
plasma and neutrinos to the gravitational field.  In this case, the comparison
of Eqs.~(5) and (6) for the long wavelength limit of the form factors with
Eq.~(29) gives
\begin{equation}
F_0=1\;,~~~F_2=-3t_L^2/2\;,~~~~~~G_1=3t_L\;,
\end{equation}
so the lengths (44) are here
\begin{equation}
d^2=6t_L^2\;,~~~~~~d'^2=0\;.
\end{equation}
Hence Eq.~(43) then gives the behaviour of $C_\ell$ for $\ell d/d_A\ll 1$ and
$\ell\geq 2$ as
\begin{equation}
\ell(\ell+1)C_\ell=8\pi^2B\left\{1-\frac{
6\ell(\ell+1)t_L^2}{d_A^2}\left[\ln\left(\frac{\bar{d}}{2d_A}\right)
+\sum_{r=1}^\ell\frac{1}{r}\right]+\dots \right\}\;,
\end{equation}
Aside from its weak dependence on $\bar{d}$, the behaviour of $C_\ell$ for $\ell
d/d_A\ll 1$ is independent of the baryon density, in agreement with  more
accurate computer calculations[8].  We can't calculate the length $\bar{d}$
without a model that would give the form factors at all wave numbers, but
$\bar{d}$  is expected to be roughly of order $d_H$, and since $d_A/d_H$ is
large the logarithm is not sensitive to the precise value of $\bar{d}$.  If for
instance we take $\bar{d}=\sqrt{3}t_L=d_A/58.5$ (the acoustic horizon at last
scattering for $\Omega_M=0.4$, $\Omega_V=0.6$, and $\Omega_B=0$) then the
quantity $\ell(\ell+1)C_\ell/8\pi^2B$ rises from unity when extrapolated to
$\ell= 0$ to
1.044 at $\ell=5$, and to 1.118 at $\ell=10$, which is probably the highest
value of $\ell$ for which the approximations leading to Eq.~(47) are reliable.

For $\ell$ of the order of $d_A/d_H$ the coefficients $C_\ell$ can be calculated
under the simplifying assumptions of this section by using the form factors
given by Eqs.~(8) and (9) in Eq.~(27).  The damping length is given in reference
1 as
\begin{equation}
d_D^2\equiv {\cal D}_L^2+\Delta{\cal D}_L^2\simeq
0.029\,t_L^2\left(\frac{8}{15(1+\xi)}+\frac{\xi^2}{2(1+\xi)^2}\right)+0.0025\,
d_H^2\;.
\end{equation}
Our results for $C_\ell$ at and below the first Doppler peak are not sensitive
to $d_D$.  We will
simplify our calculations  here by dropping the terms in Eqs.~(8) and (9)
that are proportional to the ratio $\xi/k^2t_L^2$, on the grounds that these
terms are not very different from corrections to the WKB approximation that are
not included either.  (At the first Doppler peak $\xi/k^2t_L^2$ increases with
$\xi$ and hence with $\Omega_Bh^2$, and for
$\Omega_Bh^2=0.03$ it has the value $0.20$.  But to be honest, the real reason
for dropping these terms is that they spoil the agreement of our results for the
height of the first Doppler peak with more accurate numerical calculations.)
The results obtained now depend
critically on the baryon density parameter $\xi\simeq 27\,\Omega_Bh^2$, and are
shown in Figure 1 for values of $\Omega_Bh^2$ ranging from zero to 0.03.
\begin{figure}
\begin{center}
\leavevmode \epsfbox{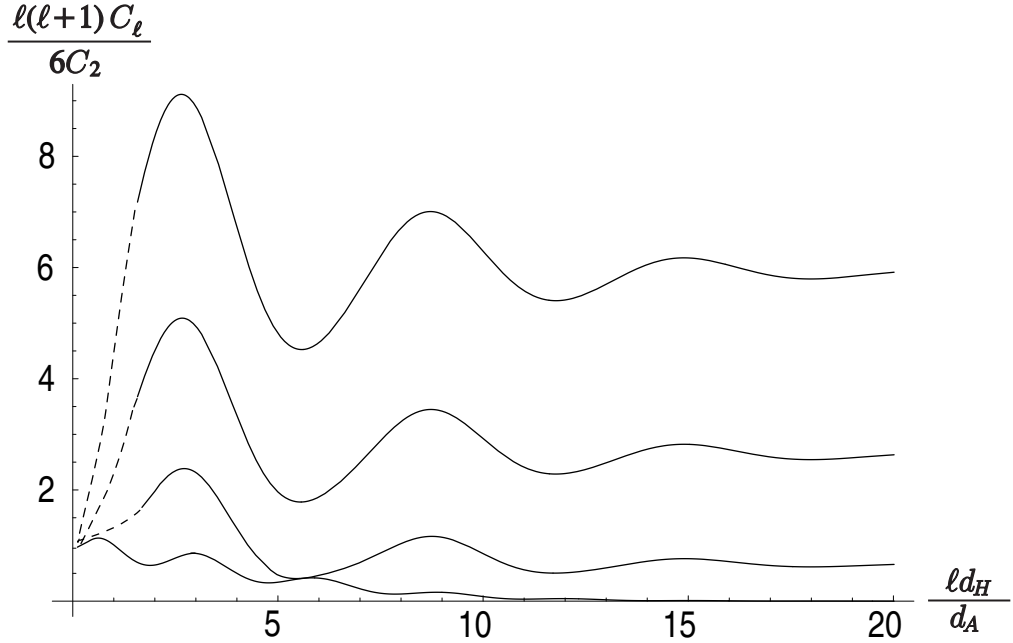}
\end{center}
\caption{Plots of the ratio of the multipole strength parameter
$\ell(\ell+1)C_\ell$ to its value at small $\ell$, versus $\ell
d_H/d_A$, where $d_H$ is the horizon size at the time of last
scattering and $d_A$  is the angular diameter distance of the
surface of last scattering.  The curves are for $\Omega_Bh^2$
ranging (from top to bottom) over the values 0.03, 0.02, 0.01, and
0, corresponding to $\xi$ taking the values 0.81, 0.54, 0.27, and
0. The solid curves are calculated using the WKB approximation;
dashed lines indicate an extrapolation to the known value at small
$\ell d_H/d_A$.  These results are independent of the parameters
$H_0$, $\Omega_\Lambda$, and $\Omega_M$.}
\end{figure}

For
$\Omega_B=0$ (in which case the WKB approximation is not needed, so that
Eq.~(27) should give $C_\ell$ down to values of $\ell$ of order two) the
behavior of $C_\ell$ is nothing like what is observed:
$\ell(\ell+1)C_\ell/8\pi^2B$ rises from unity to 1.1 at a `zeroth Doppler peak'
at $\ell d_H/d_A\simeq 0.45$ (due to the maximum in the Doppler form factor
$G(k)$ at $kd_H=\pi/2$), then dips to 0.7 at $\ell d_H/d_A\simeq 1.6$, and then
rises again to a first Doppler peak at $\ell d_H/d_A\simeq 2.83$.

For
$\Omega_Bh^2\geq 0.01$ the behavior of $C_\ell$ within the range of validity of
the WKB approximation is much more like what is observed: $\ell(\ell+1)C_\ell$
rises monotonically  to a first Doppler peak at $\ell d_H/d_A$ very roughly of
order
$\pi$ (though actually around 2.6).  There is another clear peak at $\ell \simeq
8.7
\,d_H/d_A$, presumably arising from the peak in $F(k)$ at $k=3\pi/d_H$.  The
weaker peaks in $\ell(\ell+1)C_\ell$ arising from peaks in $F(k)$ near even
values of $kd_H/\pi$ are absent here, presumably because of our neglect of the
contribution of radiation and neutrinos to the gravitational field.  Another
difference between the curves of Figure 1 and more accurate computer
calculations is that, again because we neglect  the contribution of radiation
and neutrinos to the gravitational field, our results do not show the fall-off
of $\ell(\ell+1)C_\ell$ at large $\ell$ associated with the fall-off of the
familiar transfer function $T(k)$  at large $k$.

The values of the position $\ell_1 d_H/d_A$ of the first peak and the ratio of
its
height $\ell_1(\ell_1+1)C_{\ell_1}$ to the value $8\pi^2B \simeq 6C_2$ for small
$\ell$ are given for various baryon densities  in Table 1.  These results are
independent of other parameters.  In the last two columns of Table 1 we also
give values
of
$d_A/d_H$ for $\Omega_M=0.3$ and $\Omega_\Lambda=0.7$, and the corresponding
results for the multipole number $\ell_1$ of the first Doppler peak.  In
calculating the horizon at last scattering $d_H$ we have now (somewhat
inconsistently) taken  into account the effect of photons and three flavors of
neutrinos and antineutrinos on the expansion rate, which gives
\begin{eqnarray}
d_H&\equiv &\frac{a(t_L)}{\sqrt{3}}\int_0^{t_L}\frac{dt}{a(t)\sqrt{1+R(t)}}
\nonumber\\&=&\frac{2}{H_0(1+z_L)^{3/2}\sqrt{3\xi\Omega_M}}\ln\left(\frac{\sqrt{
1+\xi }+\sqrt{\xi(1+\lambda)}}{1+\sqrt{\xi \lambda}}\right)\;,
\end{eqnarray}
where $\lambda=0.047/\Omega_M h^2$ is the ratio of photon and neutrino energy
density to dark matter energy density at the time of last scattering, and $d_A$
is given by Eq.~(2).  In calculating the values of $d_A/d_H$  in the table we
have taken $\Omega_M h^2=0.15$.

\begin{table}
\caption{Location $\ell_1 d_A/d_H$ of the first Doppler peak and height of the
peak in $\ell(\ell+1)C_\ell$ relative to its value $8\pi^2B\simeq 6C_2$ for
$\ell$ extrapolated to zero for various values of the baryon density parameter.
These
results, and the curves in Figure 1,  are independent of the values of $H_0$,
$\Omega_\Lambda$, and (within our approximations) $\Omega_M$ and.  The
last two columns give the values of $d_A/d_H$ and $\ell_1$ for $\Omega_M=0.3$,
$\Omega_\Lambda=0.7$, with $d_H$ calculated taking into account the contribution
of photons and neutrinos to the expansion rate, and using $\Omega_Mh^2=0.15$.}
\vspace*{6pt}
\centering
\begin{tabular}{||c|c|c|c||c|c||}\hline
$\Omega_Bh^2$ & $\xi$ & $\ell_1d_H/d_A$ & $\ell_1(\ell_1+1)C_{\ell_1}/6C_2$ &
$d_A/d_H$ & $\ell_1$\\ \hline
0 & 0 & 2.83 & 0.863 & 91.7 & 260 \\
0.01 & 0.27 & 2.65 & 2.34 & 87.1 & 231 \\
0.02 & 0.54 & 2.60 & 5.09 & 83.6  & 217\\
0.03 & 0.81 & 2.58 & 9.115 & 79.7  &  206 \\ \hline
\end{tabular}
\end{table}

We see from Table 1 that the position of the first Doppler peak does not depend
strongly on $\Omega_Bh^2$, while its height is a sensitive function of
$\Omega_Bh^2$.
For $\Omega_Bh^2$ between 0.02 and 0.03 the height and position are in fair
agreement with what is observed, though of
course the serious comparison of theory with observation relies on more
accurate computer calculations.  The qualitative results obtained here  suggest
that if one were to rely on a single feature of the plot of
$\ell(\ell+1)C_\ell$ versus $\ell$ to measure  $\Omega_Bh^2$, then the ratio
of the the height of the first Doppler peak to the value for
lower $\ell$ values studied by the COBE satellite would be more
useful than the ratio of the heights of the first and
second Doppler peaks, which relies on less precise data, depends on
complicated damping effects, and is more sensitive to
other parameters, such as $\Omega_Mh^2$ and the rate of
change, if any, of the vacuum energy.  Of course, for high
precision one must use the whole plot of $\ell(\ell+1)C_\ell$
versus $\ell$ to measure all these parameters together.

\begin{center}
{\bf  ACKNOWLEDGMENTS}
\end{center}
\nopagebreak
I am grateful for helpful correspondence with E. Bertschinger, J. R. Bond, L. P.
Grishchuk, and
M.
White. This research was supported in part by the Robert A. Welch Foundation and
NSF Grants PHY-0071512 and PHY-9511632.

\begin{center}
{\bf REFERENCES}
\end{center}

\begin{enumerate}

\item S. Weinberg, "Fluctuations in the Cosmic Microwave Background I: Form
Factors and their Calculation in Synchronous Gauge," UTTG 03-01,
astro-ph/0103279.

\item L. P. Grishchuk and J. Martin, Phys. Rev. D {\bf 56}, 1924 (1997).

\item J. R. Bond, ``Theory and Observations of the Cosmic Background
Radiation,'' in {\em Cosmology and Large Scale Structure}, eds. R. Schaeffer, J.
Silk, M. Spiro and J. Zinn-Justin (Elsevier, 1996), Section 5.1.3.  Our result
(26) can also be derived with somewhat more trouble by using Bond's results on
the asymptotic behavior of
averaged products of spherical Bessel functions.

\item This method had been used previously to obtain Eq.~(21) with only a scalar
form-factor $F(k)$, as for instance by J. R. Bond and G. Efstathiou, Mon.
Not. R. Astr. Soc. {\bf 226}, 655 (1987), Eq.~(4.19), but not as far as I
know with the inclusion of the dipole form-factor $G(k)$.

\item I. S. Gradshteyn and I. M. Ryzhik, {\em Table of Integrals, Series, and
Products} (Academic Press, New York, 1980), \#6.671.2.

\item I. S. Gradshteyn and I. M. Ryzhik, {\em ibid.}, \#6.574.2.

\item G. 't Hooft and M. Veltman, Nucl. Phys. {\bf B44}, 189 (1972).

\item E. F. Bunn and M. White,  Ap. J. {\bf 480}, 6 (1997).

\end{enumerate}
\end{document}